\documentclass[aip, jcp, floatfix, reprint]{revtex4-1}

\usepackage{amsmath}
\usepackage{graphicx}
\usepackage{dcolumn}
\usepackage[table]{xcolor}
\usepackage[version=3]{mhchem}
\usepackage{transparent}
\usepackage{gensymb}
\usepackage{hyperref}
\usepackage{etoolbox}

\apptocmd{\thebibliography}{\raggedright}{}{}

\def\maxfloatwidth{%
  \ifdim\columnwidth>246.0pt
  300.0pt  \else
  \columnwidth
  \fi
}

\newcommand{\trm}[1]{\textrm{#1}}

\newcommand{\etal}{\emph{et al.}}
\newcommand{\iceh}{ice I\(_{\trm{h}}\)}

\begin{document}

\date{\today}

\title{Benchmarking the performance of Density Functional Theory and
  Point Charge Force Fields in their Description of sI Methane Hydrate
  against Diffusion Monte Carlo}

\author{Stephen J. Cox} 
\affiliation{Thomas Young Centre and London Centre for Nanotechnology,
  17--19 Gordon Street, London WC1H 0AH, U.K.}
\affiliation{Department of Chemistry, University College London, 20
  Gordon Street, London, WC1H 0AJ, U.K.}

\author{Michael D. Towler} 
\affiliation{Department of Earth Sciences, University College London
  Gower Street, London WC1E 6BT, U.K.}
\affiliation{Theory of Condensed Matter Group, Cavendish Laboratory,
  University of Cambridge, J.J.~Thomson Avenue, Cambridge CB3 0HE,
  U.K.}

\author{Dario Alf\`{e}}
\affiliation{Thomas Young Centre and London Centre for Nanotechnology,
  17--19 Gordon Street, London WC1H 0AH, U.K.}
\affiliation{Department of Earth Sciences, University College London
  Gower Street, London WC1E 6BT, U.K.}

\author{Angelos Michaelides} 
\email{angelos.michaelides@ucl.ac.uk}
\affiliation{Thomas Young Centre and London Centre for Nanotechnology,
  17--19 Gordon Street, London WC1H 0AH, U.K.}
\affiliation{Department of Chemistry, University College London, 20
  Gordon Street, London, WC1H 0AJ, U.K.}

\begin{abstract}

High quality reference data from diffusion Monte Carlo calculations
are presented for bulk sI methane hydrate, a complex crystal
exhibiting both hydrogen-bond and dispersion dominated interactions.
The performance of some commonly used exchange-correlation functionals
and all-atom point charge force fields is evaluated. Our results show
that none of the exchange-correlation functionals tested are
sufficient to describe both the energetics and the structure of
methane hydrate accurately, whilst the point charge force fields
perform badly in their description of the cohesive energy but fair
well for the dissociation energetics. By comparing to \iceh, we show
that a good prediction of the volume and cohesive energies for the
hydrate relies primarily on an accurate description of the hydrogen
bonded water framework, but that to correctly predict stability of the
hydrate with respect to dissociation to \iceh{} and methane gas,
accuracy in the water-methane interaction is also required. Our
results highlight the difficulty that density functional theory faces
in describing both the hydrogen bonded water framework and the
dispersion bound methane.

\end{abstract}

\maketitle

\section{Introduction}
\label{sec:intro}

The clathrate hydrates of natural gases - crystalline compounds in
which gas is dissolved in a host framework of water molecules - are
important to a wide variety of applications across the energy and
climate sciences. For example, the fact that one volume of hydrate can
generate up to 180 volumes of gas upon dissociation at standard
temperature and pressure, whilst only 15\% of the recovered energy is
required for dissociation, means that hydrate reservoirs are a
potential untapped energy resource.\cite{clathrates} Even though there
remains uncertainty in the total amount of hydrated gas on Earth,
there is a consensus that this amount exceeds conventional gas
reserves by at least an order of
magnitude.\cite{klauda:global-hydrate-dist} Perhaps a more pressing
issue is that hydrates also pose a severe problem for flow assurance in
oil and gas pipelines: if the mixed phases of water and natural gas
are allowed to cool, hydrates may form and block the line, causing
production to stall. As readily available oil and gas reserves become
depleted, and the need for extraction from deeper reservoirs
increases, the consequences of hydrate formation are becoming more
severe. Although chemicals for inhibiting hydrate formation exist,
they have generally been found on a trial-and-error basis, with little
understanding of how they work at the molecular scale. This state of
affairs has arisen from the fact that we have little knowledge of the
fundamental mechanisms that underlie hydrate formation. Consequently,
computer simulation has been used in recent years in attempts to
improve our molecular level understanding of hydrate
formation.\cite{walsh:science, walsh:2011, molinero:blobs, trout:LSH,
  moon:faraday} It is important, therefore, to understand both the
molecular interactions present in condensed phase gas hydrates, and
the performance of current approximations used to describe these
interactions.

By far the most commonly used electronic structure method for
investigating condensed phase systems is density functional theory
(DFT) (a recent review of DFT and the current challenges it faces is
given in Ref.~\onlinecite{burke:jcp-perspective}). Despite incredible
success in its application to a wide variety of systems, DFT has a
number of limitations. Of particular relevance to gas hydrates is the
known deficiency of the local density approximation (LDA) and
generalized gradient approximation (GGA) varieties of
exchange-correlation (\(xc\)) functionals to account for van der Waals
(vdW) dispersion interactions. Incorporating an accurate description
of vdW interactions into density functional theory is a very active
research area, with recent developments including Grimme's dispersion
correction methods,\cite{DFT-D2, DFT-D3} the Tkatchenko-Scheffler
scheme\cite{TS:prl2009} and the fully self-consistent vdW-DF method of
Dion \etal{}\cite{dion:vdW-DF} and its various
derivatives.\cite{jiri:functional, jiri:functional2} For a recent
overview of these and other methods to incorporate vdW interactions
into DFT see Ref.~\onlinecite{jiri:review}. Understanding the
contribution of vdW to the bonding in solids is an important issue and
there is a need to understand the strengths and weaknesses of various
vdW-inclusive methods in order to improve the performance of
DFT. Recent work has shown that vdW-inclusive DFT methods offer a
systematic improvement over GGA functionals in describing the phase
behaviour of ice.\cite{biswajit:pressure, biswajit:pressure2} Like
ice, gas hydrates also have an extended hydrogen bonded network of
water molecules, but unlike ice, they contain cavities that gas
molecules can occupy. Natural gas hydrates therefore offer the
opportunity to test the ability of vdW-inclusive methods to
simultaneously describe both the hydrogen bonded water network and the
predominantly dispersion bound water-gas interaction. As well as DFT,
force fields (FFs) are often used to investigate gas
hydrates,\cite{walsh:science, walsh:2011, homogeneous-unrealistic,
  molinero:blobs, ojamae:clathrate, homogeneous-unrealistic}
especially when long time and length scales are required, such as in
the study of nucleation processes.

Evaluating the performance of techniques such as DFT or FFs requires
high-quality reference data to compare to -- something that is lacking
for gas hydrates in the condensed phase. For example, previous DFT
studies\cite{clathrate:vdw-DF:prl, Kolb:PRB} have evaluated the
performance of the chosen \(xc\) functional through comparison to
experiment or quantum chemical methods on isolated clusters. However,
various issues can arise when validating the performance of DFT to
experiment, such as temperature/pressure, non-stoichiometry and
quantum nuclear effects. Furthermore, although a source of valuable
information for understanding the nature of interactions, comparison
to isolated clusters (to which accurate quantum chemical methods are
generally limited to) does not directly tell us how DFT methods are
performing for the condensed phase. The tendency to validate FFs used
in molecular dynamics or Monte Carlo simulation against experiment is
even greater than it is for DFT. One method that has been shown to
provide accurate energies for condensed phase water systems is
diffusion Monte Carlo (DMC). DMC can be applied to a range of systems,
both isolated and periodic,\cite{needs:periodic-neon,
  mattsson:qmc-solids} has mild scaling behaviour\cite{qmc-scaling1,
  qmc-scaling2} and has rapid and automatic basis-set
convergence.\cite{dario:qmc-basis-convergence} DMC has also been shown
to favour well in comparison to CCSD(T) -- the so-called
`gold-standard' quantum chemical method -- for calculations on the
water dimer and other small water clusters.\cite{biswajit:jcp2008,
  gillan:jcp2012, gurtubay:qmc-dimer, ken-jordan:dmc-wat} It also
gives a good description of the relative energies of different ice
phases\cite{biswajit:pressure, biswajit:pressure2} and has recently
been shown to achieve sub-chemical accuracy for non-covalent
interactions in the gas phase.\cite{qmc:subchemical} We therefore have
confidence that DMC can be used to obtain accurate reference data for
periodic gas hydrate crystals. Specifically, we will use DMC to
calculate accurate data for the energetics of a methane hydrate
crystal.

In this study, we compare the performance of a number of different
\(xc\) functionals, ranging from the LDA and PBE\cite{PBE-GGA} levels
of approximation, in which vdW interactions are not accounted for, to
a variety of dispersion-corrected functionals, namely: an empirical
correction scheme from Grimme (PBE-D2);\cite{DFT-D2} the method
developed by Tkatchenko and Scheffler
(PBE-vdW\(^{\trm{TS}}\)),\cite{TS:prl2009} which like PBE-D2 involves
an explicit summation of pairwise vdW dispersion interactions over all
atom pairs, but differs in that the vdW \(C_{6}\) coefficients are
themselves functionals of the electron density; and a number of
functionals from the vdW-DF family. In particular, we consider the
original vdW-DF of Dion \etal{} and the modified versions of
Klime\v{s} \etal,\cite{jiri:functional, jiri:functional2} in which the
exchange functional is changed from that of revPBE, to `optPBE',
`optB88' and `optB86b'. These modified versions of vdW-DF have been
shown to offer good performance for a wide range of
systems.\cite{jiri:functional, jiri:functional2, javi:wetting,
  mittendorfer:graphene-on-Ni} Throughout the rest of the paper, the
original vdW-DF of Dion \etal{} will be referred to as `revPBE-vdW'
with the term `vdW-DF' used when referring to the class of
functionals. We also present results using the OPLS-AA\cite{opls-aa}
potential for methane and the TIP4P-2005\cite{vega:tip4p-2005} and
TIP4P-ICE\cite{vega:tip4p-ice} potentials for water. Details of these
FFs are given in the Supporting Information (SI),\footnote{See
  Supplementary Material at
  \protect\url{http://dx.doi.org/10.1063/1.4871873} for details of the
  force fields investigated and results of \iceh.}  but key features
of these potentials are that they are all atomic, point charge and
have Lennard-Jones sites located on the carbon and oxygen atoms. The
TIP4P-2005 and TIP4P-ICE potentials are rigid, whereas the OPLS-AA
potential is flexible. We have also investigated the two water
potentials in combination with a number of different methane
potentials, but as these yield similar results to OPLS-AA, they have
been omitted from the main article for clarity and are included in the
SI. Although this is clearly not an exhaustive list of possible \(xc\)
functionals and FFs available, our test set nevertheless is adequate
to highlight the main strengths and weaknesses of these types of
methods in describing hydrogen-bond plus dispersion bound systems such
as methane hydrate.

In the following sections, we will compare the results of the above
mentioned \(xc\) correlation functionals and force fields to DMC in
their prediction of the bulk properties of sI methane hydrate. We will
specifically look at the cohesive energy of the hydrate crystal, the
binding energy of the methane to the water framework and the
dissociation energy of the hydrate crystal to \iceh{} and methane
vapour. We will see that none of the methods give a particularly
satisfactory description of bulk sI methane hydrate and that in
instances of apparent agreement, this is due to a fortuitous
cancellation of errors.

\section{Computational Setup}
\label{sec:setup}

DFT calculations were performed using {\scriptsize VASP
  5.3.2},\cite{vasp1,vasp3,vasp4} a periodic plane-wave basis set
code.\footnote{For the vdW\(^{\trm{TS}}\) calculations, {\scriptsize
    VASP 5.3.3} was used.}  Calculations with the vdW-DFs have been
carried out self-consistently using the scheme of Rom\'{a}n-P\'{e}rez
and Soler,\cite{perez&soler:implementation} as implemented in
{\scriptsize VASP} by Klime\v{s} \etal.\cite{jiri:functional2}
Projector-augmented-waves (PAW) potentials\cite{vasp5} have been used,
with LDA-based potentials used for the LDA calculations and PBE-based
potentials used for all other calculations. All results reported here
used the `standard' PAW potentials supplied with {\scriptsize VASP}
and a plane-wave cutoff of 600~eV (these PAWs have been optimised for
a plane-wave basis cutoff \(\ge 400\)~eV). A \(\Gamma\)-centred
\(2\times2\times2\) Monkhorst-Pack \(k\)-point
mesh\cite{MonkhorstPack} per unit cell was used for calculations of
bulk sI methane hydrate, whereas calculations concerning isolated
molecules were performed at the \(\Gamma\)-point only, in a cubic
simulation cell of volume \(20\times20\times20\)~\AA\(^{3}\). The
structures for bulk sI methane hydrate were taken from the work of
Lenz and Ojm\"{a}e\cite{ojamae:clathrate} and optimised using the
conjugate gradient geometry optimiser until forces on all atoms were
below 0.02~eV/\AA. Wave functions were converged to within
\(1\times10^{-8}\)~eV. For calculations concerning \iceh, we have made
use of the same proton-ordered 12 molecule \iceh{} unit cell
structures as those in Ref.~\onlinecite{biswajit:pressure2}, with a
\(\Gamma\)-centred \(2\times2\times2\) Monkhorst-Pack \(k\)-point mesh
per unit cell used. All other settings were identical to those used
for the hydrate calculations.

All quantum Monte Carlo calculations were performed using version
2.12.1 of the Cambridge {\scriptsize CASINO} code.\cite{qmc-scaling2}
DMC simulations for 178-atom simulation cells were performed using
conventional Slater-Jastrow trial wave functions with a Jastrow factor
containing electron-nucleus, electron-electron, and electron-nucleus
electron terms,\cite{jas} each of which depends on variational
parameters determined by a combination of variance- and
energy-minimization. The orbitals in the determinantal part of the
trial wave function were generated from DFT calculations performed by
the PWSCF component of the Quantum Espresso package;\cite{PWSCF} these
\(\Gamma\)-point DFT calculations were done using the PBE \(xc\)
functional and a 300 Ry (4082 eV) plane-wave cutoff. The same
structures from Lenz and Ojam\"{a}e used for the {\scriptsize VASP}
calculations were first optimised with these settings in the PWSCF
component of the Quantum Espresso package. As is standard practice,
the plane-wave orbitals were re-expressed in
B-splines\cite{dario:qmc-basis-convergence} for the DMC simulations.
Dirac-Fock pseudopotentials specifically developed for use in QMC were
used.\cite{trial-needs1, trial-needs2} Although in principle
pseudopotentials for hydrogen are not required, this would imply using
e.g. Gaussian basis sets to construct the trial wavefunctions, the
completeness of which is difficult to establish in a systematic
way. We prefer to use plane waves to achieve full, automatic and
unbiased basis set convergence. The quality of the hydrogen
pseudopotential is supported in
Refs.~\onlinecite{jie:benzene1,jie:benzene2,gillan:jcp2012} where
agreement is to within 3~meV/\ce{H2O} of CCSD(T) calculations and
Ref.~\onlinecite{biswajit:pressure}, where agreement is to within
5~meV/\ce{H2O} of experiment. Coulomb finite size effects were
accounted for using the `structure factor' method described in
Refs.~\onlinecite{qmc-coulomb1} and~\onlinecite{qmc-coulomb2} (though
we could equally well have used the Modified Periodic Coulomb (MPC)
interaction defined in Refs~\onlinecite{mpc1} and~\onlinecite{mpc2},
which we checked gave essentially the same results).

FF calculations were performed using the {\scriptsize GROMACS 4.5.5}
simulation package.\cite{gromacs4} Long range electrostatics were
treated with the particle-mesh Ewald method\cite{pme1, pme2} with a
grid spacing of 1~\AA{} used for the fast Fourier transform (fourth
order interpolation was also used) and a real space cut-off of
9~\AA. Lennard-Jones interactions were truncated after 9~\AA{} with
tail corrections applied. The calculations were also performed without
the tail corrections and results from these have been included in the
SI (any effect of the tail corrections does not alter the conclusions
presented in the main paper). The L-BFGS algorithm\cite{l-bfgs1,
  l-bfgs2} was used to optimise the geometries, with the SETTLES
algorithm\cite{settle} used to constrain the water geometry. All
geometries were converged to within \(1.05 \times 10^{-6}\)~eV/\AA.

\section{Results and Discussion}
\label{sec:results}

\begin{figure}[htb]
  \centering
  \includegraphics[width=0.95\linewidth]{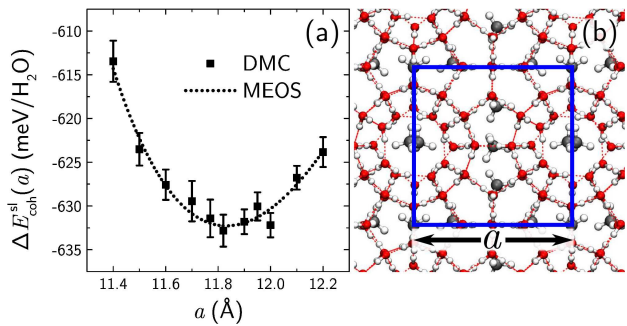}
  \caption{\textbf{ (a) Variation of the DMC cohesive energy of bulk
      sI methane hydrate with lattice constant.} The cohesive energy
    is defined by Equation~\protect\ref{eqn:cohesive}. The bars on
    each data point indicate a one standard deviation estimate of the
    error. From the fit to Murnaghan's equation of state (MEOS), the
    equilibrium lattice constant and cohesive energy are estimated to
    be \(11.83 \pm 0.02\)~\AA{} and \(-632 \pm 1\)~meV/\ce{H2O},
    respectively. \textbf{(b) Bulk sI methane hydrate crystal
      structure.} The blue box bounds the unit cell. The atoms are
    coloured as: grey, carbon; red, oxygen; and white, hydrogen. The
    dashed red lines outline the hydrogen bonded water framework.}
  \label{fig:dmc-eos}
\end{figure}

Gas hydrates come in three main crystal forms - structures I, II and H
(sI, sII and sH, respectively). Methane hydrate is generally found in
the sI form, although the sII and sH forms have been reported under
very high pressure (above 250~MPa and \emph{ca.} 1~GPa
respectively).\cite{Me-press-trans:pnas, Me-press-trans:jpcsolids} In
the sI hydrate, the water molecules form a hydrogen bonded network
that gives rise to two types of cavities: a twelve-sided pentagonal
dodecahedron (often denoted as \(5^{12}\)); and a 14-sided
tetrakaidecahedron (denoted as \(5^{12}6^{2}\), owing to the fact that
it consists of 12 pentagonal and 2 hexagonal faces). In stoichiometric
sI hydrate, the methane molecules singly occupy each cavity. The cubic
unit cell consists of two \(5^{12}\) and six \(5^{12}6^{2}\) cages and
has the chemical formula \(46\ce{H2O}\cdot 8\ce{CH4}\). The sI methane
hydrate structure is shown in Fig.~\ref{fig:dmc-eos}(b) and a
comprehensive overview of the sI, sII and sH hydrate structures can be
found in Ref.~\onlinecite{clathrates}.

We have computed the cohesive energy per water molecule:
\begin{equation}
  \label{eqn:cohesive}
  \Delta E^{\trm{sI}}_{\trm{coh}}(a) = \frac{E_{\trm{sI}}(a) - 46E_{\ce{H2O}} - 8E_{\ce{CH4}}}{46}
\end{equation}
\noindent of the bulk sI methane hydrate unit cell for a variety of
unit cell volumes, maintaining a cubic simulation cell. In
Eq.~\ref{eqn:cohesive}, \(E_{\trm{sI}}(a)\) is the total energy of
bulk sI methane hydrate with lattice constant \(a\), whilst
\(E_{\trm{H\(_{2}\)O}}\) and \(E_{\trm{CH}_{4}}\) are energies of the
isolated water and methane molecules, respectively. We did this first
using DMC. By fitting \(\Delta E^{\trm{sI}}_{\trm{coh}}(a)\) to
Murnaghan's equation of state,\cite{Murnaghan15091944, FuHu:MEOS} we
are able to determine the equilibrium lattice constant \(a_{0}\) and
cohesive energy \(\Delta E^{\trm{sI}}_{\trm{coh}}(a_{0})\). These
results are presented in Fig.~\ref{fig:dmc-eos}(a), where we can see
that the equilibrium lattice constant is estimated to be \(11.83 \pm
0.02\)~\AA{} and the cohesive energy is \(-632 \pm
1\)~meV/\ce{H2O}. The DMC lattice constant compares well to the low
temperature neutron scattering data of Davidson
\etal{}\cite{davidson:nature-clathrate} (\(11.77 \pm 0.01\)~\AA{},
\ce{CH4}/\ce{D2O} at 5.2~K) and Gutt \etal{}\cite{gutt:clathrate}
(\(11.821 \pm 0.001\)~\AA{}, \ce{CD4}/\ce{D2O} at 2~K).

\begin{figure}[htb]
  \centering
  \includegraphics[width=0.85\linewidth]{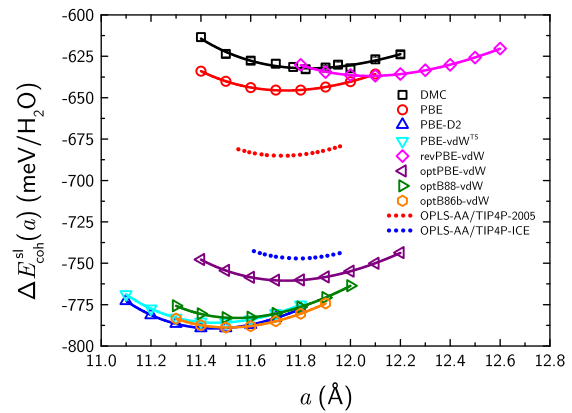}
  \caption{\textbf{Variation of the cohesive energy of bulk sI methane
      hydrate with lattice constant.} The cohesive energy is defined
    by Equation~\protect\ref{eqn:cohesive}. Symbols represent
    calculated values using DFT (empty squares show DMC data), whereas
    the solid lines show a fit to Murnaghan's equation of state. For
    the DMC data, the error bars are smaller than the size of the
    symbols. Results using the OPLS-AA force field for methane in
    combination with the TIP4P-2005 and TIP4P-ICE water potentials are
    also shown (fit to Murnaghan's equation of state only). The
    results for LDA, which has a cohesive energy of
    \(-1178\)~meV/\ce{H2O} and equilibrium lattice constant of
    \(10.933\)~\AA, have been omitted for clarity.}
  \label{fig:binding-curves}
\end{figure}

We have also computed the variation of the cohesive energy with
lattice constant for each of the DFT \(xc\) functionals and FFs
discussed in Section~\ref{sec:intro}. In these calculations, all atoms
were allowed to relax independently (with the constraint of rigid
water molecules for the FFs). The results of these calculations are
presented in Figure~\ref{fig:binding-curves} and
Table~\ref{tab:results} (the LDA results have been excluded from
Fig.~\ref{fig:binding-curves} for clarity). Although all of the
examined DFT \(xc\) functionals overbind the hydrate crystal, there is
considerable variety amongst the DFT results, with optPBE-, optB88-,
optB86b-vdW, PBE-D2 and PBE-vdW\(^{\trm{TS}}\) significantly
overbinding the hydrate crystal, whilst PBE and revPBE-vdW yield
cohesive energies in better agreement with DMC. Despite having the
best agreement with DMC for the cohesive energy (within 1\%),
revPBE-vdW does, however, predict a lattice constant that is
1.9--2.3\% too large. On the other hand, although it significantly
overbinds the crystal, optPBE-vdW predicts a structure in decent
agreement with DMC (0.6--0.9\% too small), whereas optB88- and
optB86b-vdW yield lattice constants that are too short by 2.3--2.6\%
and 2.5--2.9\%, respectively. PBE-D2 and PBE-vdW\(^{\trm{TS}}\) also
strongly overbind the hydrate crystal and predict lattice constants
that are too small by 3.0\% or worse. What is perhaps surprising is
that PBE, which fails to account for vdW interactions entirely, is
yielding reasonable results not only for the structure (0.6--0.9\%
smaller than DMC), but also for the energetics. In fact, not only does
PBE predict an equilibrium cohesive energy in reasonable agreement
with the DMC result, it actually slightly overbinds the crystal by
2.1--2.4\%. The force fields, OPLS-AA/TIP4P-2005 and
OPLS-AA/TIP4P-ICE, overbind by 8.2-8.6\% and 18.0-18.4~\%
respectively, although their predicted structures are in decent
agreement with the reference data, with their predicted lattice
constants differing from DMC by less than \(1.0\%\).

\begin{figure}[htb]
  \centering
  \includegraphics[width=0.85\linewidth]{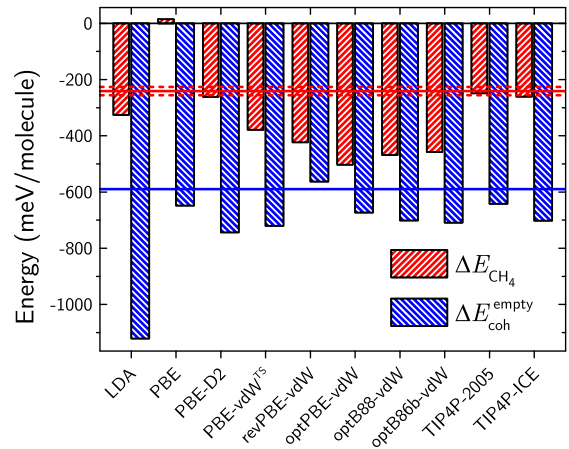}
  \caption{\textbf{Binding energy of methane to the empty hydrate
      \(\Delta E_{\ce{CH4}}\) and formation energy of the empty
      hydrate \(\Delta E^{\trm{empty}}_{\trm{coh}}\)}. The horizontal
    red and blue lines show the DMC values for \(\Delta E_{\ce{CH4}}\)
    and \(\Delta E^{\trm{empty}}_{\trm{coh}}\) respectively (dashed
    lines show the associated statistical uncertainty - not visible
    for \(\Delta E^{\trm{empty}}_{\trm{coh}}\)). Apart from PBE-D2,
    all dispersion-corrected density functionals severely overbind
    methane to the empty hydrate. Similarly, all density functionals
    overbind the water framework, with the exception of
    revPBE-vdW. PBE, which does not account for vdW interactions,
    fails to predict methane binding to the empty hydrate
    structure. The force fields yield good values for \(\Delta
    E_{\ce{CH4}}\), but like the DFT methods, they overbind the water
    framework.}
  \label{fig:energy-decomp}
\end{figure}

From the results for \(\Delta E^{\trm{sI}}_{\trm{coh}}(a)\) presented
in Fig.~\ref{fig:binding-curves} it would be tempting to conclude that
PBE gives a satisfactory description of bulk sI methane hydrate. Given
the well known problem that GGA functionals do not account for
dispersion interactions, however, the fact that PBE slightly overbinds
the hydrate seems almost paradoxical. Furthermore, the overly
repulsive nature of revPBE exchange at short separations has been
shown to lead to lattice constants that are too long and cohesive
energies that are too weak in hydrogen bonded systems such as
ice.\cite{biswajit:pressure2} Indeed, we do obtain a lattice constant
with revPBE-vdW that is 1.9--2.3\% too large, but why then, is the
cohesive energy for sI methane hydrate slightly too strong with this
functional? To better understand these results, we have decomposed the
total cohesive energy into contributions arising from the methane
binding to the empty hydrate:
\begin{equation}
  \label{eqn:ch4-binding}
  \Delta E_{\ce{CH4}} = \frac{E_{\trm{sI}}(a_{0}) - E_{\trm{empty}}(a_{0}) - 8E_{\ce{CH4}}}{8}
\end{equation}
\noindent and the cohesive energy of the empty hydrate:
\begin{equation}
  \label{eqn:empty-formation}
  \Delta E^{\trm{empty}}_{\trm{coh}} = \frac{E_{\trm{empty}}(a_{0}) - 46E_{\ce{H2O}}}{46}
\end{equation}
\noindent where \(E_{\trm{empty}}(a_{0})\) is the energy of the
hydrate unit cell with no methane present, calculated without further
relaxation of the water molecules (i.e. the water molecules are
`frozen' in the position they assume in the bulk hydrate). For DMC,
both \(\Delta E_{\ce{CH4}}\) and \(\Delta
E^{\trm{empty}}_{\trm{coh}}\) have been calculated at the experimental
lattice constant\cite{davidson:nature-clathrate} \(a =
11.77\)~\AA. These results are presented in
Fig.~\ref{fig:energy-decomp} and Table~\ref{tab:results}, with DMC
providing reference values of \(\Delta E_{\ce{CH4}} = -241 \pm
15\)~meV/\ce{CH4} and \(\Delta E_{\trm{coh}}^{\trm{empty}} = -590 \pm
2\)~meV/\ce{H2O}. The origin of PBE's seemingly good description of
bulk sI hydrate now becomes apparent: the lack of vdW interactions
means there is no binding between the methane and the water (in fact
\(\Delta E_{\ce{CH4}}\) is slightly positive), but this is compensated
for by an overbinding of the hydrogen bonded water framework. Although
the overbinding of the water framework is small on a per molecule
basis, water and methane exist in a ratio of 23:4 in the
stoichiometric hydrate, meaning that small errors in describing the
water-water interactions are much amplified compared to the apparently
larger errors in the methane binding energy. From
Fig.~\ref{fig:energy-decomp} we can also see that LDA's severe
overbinding occurs principally from its description of the water
framework (\(\Delta E^{\trm{empty,LDA}}_{\trm{coh}} - \Delta
E^{\trm{empty,DMC}}_{\trm{coh}} = -531\)~meV/\ce{H2O}), although it is
worth noting that it also overbinds the methane to the water framework
by 85~meV/\ce{CH4}. The ability of LDA to bind van der Waals systems
(such as \ce{CH4} in a \ce{H2O} cage) has been observed
before;\cite{water-on-graphene, gabriella:improved-layers} this is
known to be fortuitous because, by its nature, LDA relies on a local
description of exchange and correlation and does not account for
non-local interactions. Turning our attention to the
dispersion-corrected functionals it is clear that, with the exception
of PBE-D2, they all over-correct the neglect of vdW interactions by
the GGA functional, yielding methane binding energies that are too
strong by 138--262~meV/\ce{CH4}. It is also clear that the better
agreement of the cohesive energy obtained with revPBE-vdW compared to
the other dispersion-corrected functionals is due to an underbinding
of the water framework (consistent with results obtained for bulk
\iceh\cite{biswajit:pressure2}) that offsets a strong overbinding of
the methane. In the case of the other dispersion-corrected
functionals, as well as predicting methane binding energies that are
too exothermic, they also overbind the water framework by
83--154~meV/\ce{H2O}. The source of overbinding for the FFs occurs
almost exclusively in the water framework, with both FFs presented
here yielding good agreement for \(\Delta E_{\ce{CH4}}\).

\begin{table*}[htb]
  \centering
  \begin{tabular}{c c c c c c c}
    \hline
    \hline
     Method      &  \(\Delta E^{\trm{sI}}_{\trm{coh}}(a_0)\)     &  \(a_{0}\)    &  \(\Delta E_{\ce{CH4}}\)   &  \(\Delta E^{\trm{empty}}_{\trm{coh}}\)  &  \(\Delta E^{\trm{ice}}_{\trm{coh}}\)  &  \(\Delta E_{\trm{diss}}^{\trm{sI}\rightarrow\trm{ice}}\)  \\        
    \hline                                                                                                                                                                                                                                                                         
    LDA                  &  \(-1178\)                             &  \(10.933\)  &           \(-326\)         &   \(-1121\)                              &  \(-1136\)                             &  \(+240\)   \\                                                         
    PBE                  &  \(-646\)                              &  \(11.740\)  &           \(+15\)          &   \(-648\)                               &  \(-657\)                              &  \(-67 \)   \\                                                         
    PBE-D2               &  \(-789\)                              &  \(11.453\)  &           \(-262\)         &   \(-744\)                               &  \(-758\)                              &  \(+179\)   \\                                                         
    PBE-vdW\(^{\trm{TS}}\) &  \(-786\)                            &  \(11.461\)  &           \(-379\)         &   \(-720\)                               &  \(-737\)                              &  \(+280\)   \\                                                        
    revPBE-vdW           &  \(-637\)                              &  \(12.077\)  &           \(-423\)         &   \(-563\)                               &  \(-583\)                              &  \(+308\)   \\                                                         
    optPBE-vdW           &  \(-760\)                              &  \(11.743\)  &           \(-503\)         &   \(-673\)                               &  \(-696\)                              &  \(+369\)   \\                                                         
    optB88-vdW           &  \(-783\)                              &  \(11.542\)  &           \(-468\)         &   \(-702\)                               &  \(-725\)                              &  \(+335\)   \\                                                         
    optB86b-vdW          &  \(-789\)                              &  \(11.509\)  &           \(-458\)         &   \(-709\)                               &  \(-733\)                              &  \(+321\)   \\                                                         
    OPLS-AA/TIP4P-2005   &  \(-685\)                              &  \(11.726\)  &           \(-248\)         &   \(-642\)                               &  \(-653\)                              &  \(+184\)   \\                                                         
    OPLS-AA/TIP4P-ICE    &  \(-747\)                              &  \(11.792\)  &           \(-261\)         &   \(-702\)                               &  \(-714\)                              &  \(+190\)   \\                                                         
    \hline                                                                                                                                                                     
    DMC                  & \(-632\pm 1\)                          &  \(11.83 \pm 0.02\)  &   \(-241\pm15\)    &  \(-590\pm2\)                            &  \(-605 \pm 5\)                       &  \(+155 \pm 34\)   \\                                                         
    \hline
    \hline
  \end{tabular}
\caption{\textbf{Computed cohesive energies \(\Delta
    E^{\trm{sI}}_{\trm{coh}}(a_0)\), equilibrium lattice constants
    (\(a_{0}\)), methane binding energies to the empty hydrate
    (\(\Delta E_{\ce{CH4}}\)), empty hydrate cohesive energies
    (\(\Delta E^{\trm{empty}}_{\trm{coh}}\)), ice cohesive energies
    (\(\Delta E^{\trm{ice}}_{\trm{coh}}\)), and methane hydrate
    dissociation energies to \iceh{} and gas (\(\Delta
    E_{\trm{diss}}^{\trm{sI}\rightarrow\trm{ice}}\)).} The DMC value
  for \(\Delta E^{\trm{ice}}_{\trm{coh}}\) is taken from
  Ref.~\onlinecite{biswajit:pressure}. The unit of \(\Delta
  E^{\trm{sI}}_{\trm{coh}}(a_0)\), \(\Delta
  E^{\trm{empty}}_{\trm{coh}}\) and \(\Delta
  E^{\trm{ice}}_{\trm{coh}}\) are meV/\ce{H2O}, whilst \(\Delta
  E_{\ce{CH4}}\) and \(\Delta
  E_{\trm{diss}}^{\trm{sI}\rightarrow\trm{ice}}\) are given in
  meV/\ce{CH4}. The equilibrium lattice constant \(a_{0}\) is given in
  \AA ngstrom.}
  \label{tab:results}
\end{table*}

\begin{figure}
  \centering
  \includegraphics[width=0.95\linewidth]{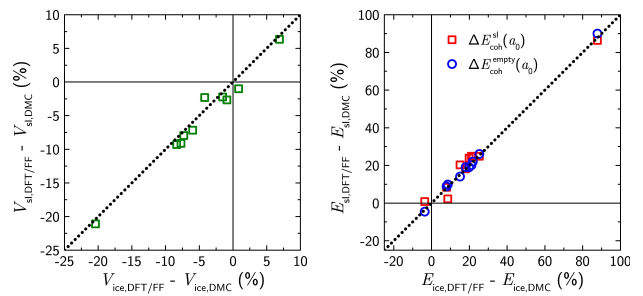}
  \caption{\textbf{Comparison of \(xc\) functional and FF performance
      for \iceh{} and sI methane hydrate.} The left panel shows the
    percentage difference from the DMC sI hydrate volume against the
    percentage difference from the DMC \iceh{} volume, for the various
    DFT \(xc\) functionals and force fields. The right panel shows the
    percentage difference from the DMC sI hydrate cohesive energies
    (\(\Delta E^{\trm{sI}}_{\trm{coh}}\) and \(\Delta
    E^{\trm{empty}}_{\trm{coh}}\)) against the percentage difference
    from the DMC cohesive energy for \iceh, again for all the \(xc\)
    functionals and FFs investigated.}
  \label{fig:comparisons}
\end{figure}

Due to the high water content of sI methane hydrate, it is convenient
to compare the performance of the \(xc\) functionals and FFs for the
hydrate to \iceh. We choose \iceh{} rather than any of the other
phases of ice due to its close structural similarity to sI hydrate at
the molecular level: the average hydrogen bond length in the hydrate
is only 1\% longer on average than in \iceh{} and the hydrate O--O--O
angles differ from the tetrahedral angles of \iceh{} by only
3.7\degree.\cite{clathrates} In the same manner that we calculated
\(\Delta E^{\trm{sI}}_{\trm{coh}}(a_{0})\) for sI methane hydrate, we
have also calculated \(\Delta E^{\trm{ice}}_{\trm{coh}}\) for our test
set of \(xc\) functionals and FFs by fitting the cohesive energy of
the bulk \iceh{} crystal to Murnaghan's equation of state (a more
comprehensive overview of the ice results is given in the SI). For
DMC, we use the value of \(\Delta E^{\trm{ice}}_{\trm{coh}}\) reported
in Ref.~\onlinecite{biswajit:pressure}. From these calculations, we
also obtain the equilibrium volume of the \iceh{} crystal. In
Fig.~\ref{fig:comparisons} we show the difference in computed volume
using DFT/FF from that using DMC for sI methane hydrate, plotted
against the same quantity for \iceh. There is a strong correlation
between the errors in computed volumes for the hydrate and \iceh,
suggesting that the primary factor in obtaining reasonable lattice
volumes for the hydrate is an accurate description of the hydrogen
bonded water framework. We also show in Fig.~\ref{fig:comparisons} the
differences in DFT/FF values for \(\Delta E^{\trm{sI}}_{\trm{coh}}\)
and \(\Delta E^{\trm{empty}}_{\trm{coh}}\) from DMC, again plotted
against the DFT/FF-DMC difference for the \iceh{} cohesive energy. As
for the volumes, we see a very strong positive correlation between the
errors in the hydrate cohesive energies and those for \iceh. In fact,
there is a near perfect correlation for the error \(\Delta
E^{\trm{empty}}_{\trm{coh}}\) and the error in \(\Delta
E^{\trm{ice}}_{\trm{coh}}\), the significance of which will become
apparent when we look at the dissociation behaviour of the hydrate to
\iceh{} and methane gas.

In comparing the cohesive energies of the DFT \(xc\) functionals and
point charge FFs to DMC, we are taking the vapour phase of both
methane and water as our reference state. More important to the phase
equilibria of gas hydrates, however, is the relative energy of the
hydrate with respect to methane gas and another condensed phase of
water, either liquid or ice.\cite{clathrates} Whilst the cancellation
of errors in \(\Delta E_{\ce{CH4}}\) and \(\Delta
E^{\trm{empty}}_{\trm{coh}}\) means that PBE has a good overall
agreement with the DMC cohesive energy, it is straightforward to
demonstrate that the error in \(\Delta E_{\ce{CH4}}\) arising from the
neglect of vdW interactions can lead to severe consequences regarding
the thermodynamic stability of sI methane hydrate. Consider the
process of sI methane hydrate dissociating to \iceh{} and methane gas:
%
  \begin{equation}
    \label{chemeqn:dissoc-to-ice}
    5.75\ce{H2O}\cdot\ce{CH4}_{\trm{(sI)}} \xrightarrow{\Delta E_{\trm{diss}}^{\trm{sI}\rightarrow\trm{ice}}} 5.75\ce{H2O}_{\trm{(ice)}} + \ce{CH4}_{\trm{(gas)}}
  \end{equation}
%
\noindent The associated energy cost \(\Delta
E_{\trm{diss}}^{\trm{sI}\rightarrow\trm{ice}}\) can be computed as
(see SI):
%
  \begin{equation}
    \label{eqn:dissoc-to-ice}
    \Delta E_{\trm{diss}}^{\trm{sI}\rightarrow\trm{ice}} = 5.75\Delta E^{\trm{ice}}_{\trm{coh}} - 5.75\Delta E^{\trm{empty}}_{\trm{coh}} - \Delta E_{\ce{CH4}}
  \end{equation}
%
The results of these calculations are presented in
Table~\ref{tab:results}. It is clear that the results for PBE are
disastrous: sI methane hydrate is unstable with respect to
dissociation to \iceh{} and methane gas by 67~meV/\ce{CH4} (i.e. it is
67~meV/\ce{CH4} exothermic). In contrast, DMC predicts dissociation to
be an endothermic process, costing \(155 \pm 34\)~meV/\ce{CH4}. We
note here that the experimental enthalpy of
dissociation\cite{handa:enthalpies} at standard temperature and
pressure is \(188 \pm 3\)~meV/\ce{CH4} suggesting that the DMC value
is reasonable.\footnote{As the experimental number is a standard
  enthalpy of dissociation we should not expect quantitative agreement
  with the DMC dissociation energy, which is a total energy
  difference. Aside from the temperature/pressure effects present in
  experiment, there is also the issue of non-stoichiometry (the
  experimental data of Handa\protect\cite{handa:enthalpies} was
  obtained for a methane occupancy of water cages of \emph{ca.} 96\%),
  which means that configurational entropy is likely to be important
  for the experimental dissociation enthalpy. This comparison is made
  simply to show that the number obtained with DMC is reasonable. In
  fact, analysis of numerous experimental data sets using the
  Clapeyron equation yields an enthalpy of dissociation of \(157 \pm
  6\)~meV/\ce{CH4} at 150~K and
  0.0564~bar.\protect\cite{anderson:enthalpies} This is arguably a
  better comparison to the zero temperature/pressure DMC calculations
  and indeed improves agreement, but one should nevertheless exercise
  caution when comparing a calculated dissociation energy to an
  experimental enthalpy.} All of the dispersion-corrected functionals
improve on the GGA functional in this respect, predicting that the
hydrate is stable with respect to ice and methane gas. PBE-D2 gives
the best agreement with DMC, followed by LDA and
PBE-vdW\(^{\trm{TS}}\), although it should be kept in mind that these
calculations have been performed at the equilibrium volume of the
\(xc\) functional used. The vdW-DFs over-stabilise the hydrate by a
factor of approximately two. Unsurprisingly, the trends in \(\Delta
E^{\trm{ice}}_{\trm{coh}}\) closely follow those of \(\Delta
E_{\ce{CH4}}\), with the errors in describing the hydrogen bonded
water network more-or-less cancelling between \(\Delta
E^{\trm{empty}}_{\trm{coh}}\) and \(\Delta E^{\trm{ice}}_{\trm{coh}}\)
(as shown in Fig.~\ref{fig:comparisons}). As such, the FFs also give
good agreement with the DMC result. The fact that the point charge FFs
predict \(\Delta E_{\trm{coh}}^{\trm{sI}}\) to be too exothermic can
be attributed to the enhanced dipole moment of the isolated water
molecules in these types of potentials,\cite{berendsen:spce,
  Guillot:water-model-review} which has been shown to lead to too high
vaporisation enthalpies of \iceh{} for the TIP4P-2005 and TIP4P-ICE
potentials.\cite{vega:ice-dipoles} Indeed, Vega and
co-workers\cite{vega:ice-dipoles} have found that it is impossible to
simulataneously fit the melting temperature of \iceh{} and the
enthalpy of vaporisation for such models. It is therefore probably
expecting too much of the rigid point charge FFs to give reasonable
results for both \(\Delta E_{\trm{coh}}^{\trm{sI}}\) and \(\Delta
E_{\trm{diss}}^{\trm{sI}\rightarrow\trm{ice}}\) whilst also
maintaining favourable densities and coexistence/melting temperatures
for the hydrate and \iceh.\cite{vega:coexist} Use of an explicitly
polarizable water potential may go some way to improving this
situation.\cite{ken-jordan:polarizable-hydrate}

\section{Conclusions}
\label{sec:concl}

We have presented high-quality DMC reference data for bulk sI methane
hydrate and evaluated the performance of several commonly used \(xc\)
functionals and point charge force fields. We have found that none of
the DFT methods tested give particularly satisfactory results. We find
that vdW forces are crucial to the stability of methane hydrate with
respect to dissociation to \iceh{} and methane gas, although the
vdW-DF flavour of \(xc\) functionals over-stabilise the hydrate by
approximately a factor of two. This effect is less severe with the
PBE-D2 and PBE-vdW\(^{\trm{TS}}\) functionals, although their
equilibrium volumes are too small compared to DMC. PBE, which neglects
dispersion interactions, incorrectly predicts that methane hydrate is
unstable with respect to dissociation to ice and methane gas. By
overbinding the hydrogen bonded water framework, PBE's poor
description of the water-methane interaction is compensated, giving a
good overall agreement with the DMC cohesive energy of the bulk
hydrate. This last point highlights the difficulty that DFT \(xc\)
functionals face in describing mixed phase systems such as gas
hydrates; in order to obtain a good overall description, it is
necessary to be able to accurately describe both the hydrogen bonded
water framework and the dispersion bound methane. We have also seen
that point-charge, all-atom force fields tend to overbind the hydrate
lattice, although their agreement with DMC for the dissociation energy
to ice and vapour, and for the structure for the bulk crystal, is
good. From our knowledge of the literature\cite{vega:ice-dipoles} on
the performance of simple point charge FFs for ice, it is unlikely
that such FFs will be able to simultaneously describe both the
cohesive energy of the hydrate crystal and the energetics of
dissociation to other condensed phase water systems.

Earlier in this article, we remarked that the 23:4 ratio of water to
methane amplified the apparently small errors in the water-water
interactions compared to the water-methane interactions. We also saw
that the high water content means that the errors in describing the
hydrate are strongly correlated to the errors in describing
\iceh. However, such a high water-methane ratio also means that there
is the possibility for significant many body interactions between the
methane and water (e.g. a single isolated \(5^{12}\) cage has 190
water-methane-water triplets). Indeed, a separate independent study
investigating the binding energy of methane to a gas phase \(5^{12}\)
cage through a many-body expansion of the total energy has found
significant contributions to the DFT error beyond those in the
two-body interactions (symmetry adapted perturbation theory
calculations also showed that the DFT methods have insufficiencies
other than those associated with the neglect of long-range dispersion
interactions).\cite{ken-jordan:MB}

Although we have not considered the effects of including exact
exchange, it is unlikely that this will significantly improve the DFT
description of sI methane hydrate. For example, using the
PBE0\cite{pbe0} results for \iceh{} from
Refs.~\onlinecite{biswajit:pressure2}, and for a methane molecule
binding to a gas phase \(5^{12}\) water cage from
Ref.~\onlinecite{ken-jordan:MB}, we predict that this hybrid \(xc\)
functional will give a reasonable prediction of the hydrate structure
(similar to PBE) but will still incorrectly destabilise the hydrate
with respect to methane gas and \iceh. Including dispersion
corrections to this functional, such as PBE0-D2 or
PBE0-vdW\(^{\trm{TS}}\), can therefore be expected to also give
similar results to PBE-D2 and PBE-vdW\(^{\trm{TS}}\). It has recently
been shown\cite{dario:jcp-manybody} that accurate DMC reference data
in combination with Gaussian approximation potentials\cite{GAP-prl}
can be used to systematically correct the `beyond two-body' errors
associated with GGA functionals for water nano-droplets and bulk
liquid water. Such an approach is also likely to be more successful in
improving the performance of DFT \(xc\) functionals for gas hydrates
compared to the pairwise additive dispersion corrections examined
here.

\section{Acknowledgements}
\label{sec:ack}

We would like to thank Dr. Biswajit Santra for sharing the structures
used in the calculations for \iceh. Professor Mike Gillan, Professor
Ken Jordan and Dr. Ben Slater are thanked for their reading of this
manuscript and useful discussions. This research used resources of the
Oak Ridge Leadership Computing Facility at the Oak Ridge National
Laboratory, which is supported by the Office of Science of the
U.S. Department of Energy (DOE) under Contract No. DEAC05-
00OR22725. We are grateful to the London Centre for Nanotechnology and
UCL Research Computing for computational resources. M.D.T. is
supported by the Engineering and Physical Sciences Research Council
though grant N. EP/K038249/1. A.M. is supported by the European
Research Council and the Royal Society through a Royal Society Wolfson
Research Merit Award.

\bibliography{bibliography}

\end{document}